\documentclass[prl,twocolumn]{revtex4}
\usepackage{amssymb}
\usepackage{graphicx}

\def\v#1{\textbf{\emph{#1}}}

\def\t#1{\widetilde{#1}}

\def\ket#1{\left|#1\right\rangle}
\def\tx#1{{\rm{#1}}}

\def\eps{\epsilon}

\def\skiip#1{}

\begin{document}
\title{Tensor-entanglement renormalization group approach
to topological phases}
\author{Zheng-Cheng Gu$^\dagger$, Michael Levin$^{\dagger\dagger}$ and  Xiao-Gang Wen$^{\dagger}$}
\affiliation{ Department of Physics, Massachusetts Institute of
Technology, Cambridge, Massachusetts 02139, USA$^{\dagger}$
\\Department of Physics, Harvard University, Cambridge,
Massachusetts 02138, USA $^{\dagger\dagger}$ }

\begin{abstract}
The tensor-entanglement renormalization group approach is applied to
Hamiltonians that realize a class of topologically ordered states --
string-net condensed states. We analyze phase transitions between
phases with and without string-net condensation. These phase
transitions change topological order without changing any
symmetries. This demonstrates that the tensor-entanglement
renormalization group approach can be used to study the phase
diagram of a quantum system with topologically ordered phases.
\end{abstract}
\maketitle

\emph{Introduction:} A mean field approach could potentially be very useful
for understanding and analyzing topological phases\cite{Wtoprev} and
topological phase
transitions.\cite{WWtran,CFW9349,SMF9945,RG0067,Wctpt,Wqoslpub,LWstrmsh} The main
challenges in developing such an approach are (a) finding a class of mean
field states that can describe topological phases and (b) finding a simple way
to calculate the physical properties, such as average energy, of these states.
The first problem can be solved with a general class of trial wave functions
known as ``tensor product states" (or alternatively ``projected entangled pair
states"). \cite{FrankPEPS2,AV0804} Indeed, one can show that tensor product
states (TPS) can describe all the string-net condensed states constructed in
\cite{LWstrnet}, and hence all non-chiral topological phases. \cite{GuString}
The second problem can be solved with the tensor-entanglement renormalization
group (TERG) method\cite{GLWtergV} which allows one to calculate correlation
functions (including average energy) of a TPS very efficiently. More
specifically, the relative error $\eps$ in a TERG calculation scales with the
computation time $T$ like $\eps\sim \e^{-\text{const.} \cdot (\ln T)^2}$ for
2D gapped systems. \cite{GLWtergV} (For comparison, the error in a variational
Monte Carlo calculation scales like $\eps \sim \e^{-\frac12 \ln T}$, if there
is no sign problem).

In this paper, we apply the TERG approach to a few models with
string-net condensation. We calculate the phase
diagram of these systems and study phase transitions from
string-net condensed states to states without string-net
condensation. These transitions are examples of continuous phase
transitions between phases with different topological orders but
the same symmetry.\cite{WWtran,CFW9349,SMF9945,RG0067,Wctpt,Wqoslpub}
As such, they are beyond the Landau symmetry breaking paradigm.
Thus the TERG approach is capable of describing phases and phase
transitions that cannot be described by Landau's symmetry breaking
theory.

\emph{$Z_2$ gauge model:} The first system that we study
is a spin-1/2 system where the spins live on links of a square lattice.
The Hamiltonian is given by
\begin{eqnarray}
\label{HZ2}
 H= U\sum_v \Big(1-\prod_{l\in v} \sigma^z_l \Big)
-g \sum_p \prod_{l\in p} \sigma^x_l
-J\sum_l \si^z_l,
\end{eqnarray}
Here $\prod_{l\in p}\sigma^x_l$ is the product of the four $\sigma^x_l$ around
a square $p$ and $\sum_p$ sums over all the squares.  $\prod_{l\in
v}\sigma^z_l$ is the product of the four $\sigma^z_l$ around a vertex $v$ and
$\sum_v$ sums over all the vertices.  $\sum_l$ sums over all links.  We will
assume that $U=\infty$ and study the quantum phases of the above system as we
change $g$ and $J$. We will assume $J>0$ and $g>0$.

When $J=0$, \eqn{HZ2} is exactly soluble.\cite{K032} To understand the exact
ground state in the string language,\cite{Wen04}  we interpret the $\sigma^z =
-1$ and $\sigma^z = 1$ states on a single link as the presence or absence of a
string.  
The appropriate low energy Hilbert space in large $U$ limit is made of
closed string states that satisfy $\prod_{i\in v} \sigma^z_i=1$ at every
vertex. The ground state is simply an equal weight superposition of all closed
string states
$\ket{\Psi_{Z_2}}=\sum_{X \rm{closed}} \ket{X}$,
which is called a string-net condensed state.

When $g=0$, the ground state is the spin polarized state with no
down spins and no closed strings.  The above two states have the
same symmetry.  But due to the non-trivial topological order in the
string-net condensed state, the two states belong to two different
quantum phases. We would like to use the TERG approach to study the
phase transition between the above two states with different
topological orders.

We would like to mention that the low energy effective theory of 
\eqn{HZ2}
is $Z_2$ gauge theory.\cite{RS9173,Wsrvb,K032}
The transition between the string-net condensed and non condensed phases
is nothing but the transition between the deconfined and confined phases
of $Z_2$ gauge theory.

One way to study such a phase transition is to introduce a
variational wave function
\begin{equation}
\label{Psial}
\ket{\Psi_w}=\sum_{X \rm{closed}} w^{L_X}\ket{X},
\end{equation}
where $L_X$ is the number of links on the string $X$.  When $w=1$,
$\ket{\Psi_w}$ becomes the string condensed state $\ket{\Psi_{Z_2}}$. When
$w=0$, $\ket{\Psi_w}$ is the state with all spins in up direction which does
not contain any strings.  We see that $w$ describes the string tension.
$\ket{\Psi_{w=1}}$ is the ground state of \eqn{HZ2} when $J=0$ and
$\ket{\Psi_{w=0}}$ is the ground state of \eqn{HZ2} when $g=0$.

Since $\ket{\Psi_{w=1}}$ and $\ket{\Psi_{w=0}}$ have the same
symmetry, one might expect that as we change $g/J$,
$\ket{\Psi_{w=1}}$ would change into $\ket{\Psi_{w=0}}$ smoothly and
the ground state energy of \eqn{HZ2} would be a smooth function of
$g/J$, implying that there was no quantum phase transition. In fact,
we will see below that the ground state energy of \eqn{HZ2} is
not a smooth function of $g/J$ indicating that there is quantum
phase transition at a critical value $(g/J)_c$.

\begin{figure}
\begin{center}
\includegraphics[scale=0.5]
{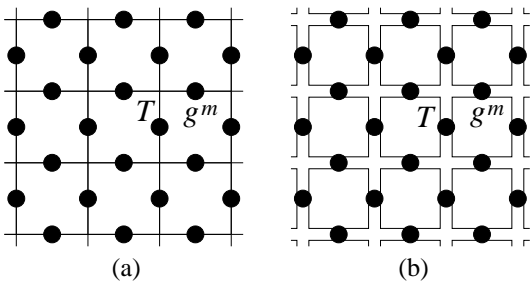}
\end{center}
\caption{
(a) The spin-1/2 model \eqn{HZ2} on links of a square lattice.
The dots represent the
physical spin states which are labeled by $m=0,1$.  The above graph can also
be viewed as a tensor-network where each dot represents a rank-3 tensor $g$
and each vertex represents a rank-4 tensor $T$.  The two legs of a dot
represent the $\alpha$ and $\beta$ indices in the rank-3 tensor
$g^m_{\alpha\beta}$.  The four legs of a vertex represent the four internal
indices in the rank-4 tensor $T_{\alpha\beta\gamma\lambda}$.  The indices on
the connected links are summed over which define the tensor trace tTr.
(b)
A tensor-network of the double-line tensors, where each dot represents a
double-line tensor $g$ and each vertex represents a double-line tensor $T$.
The four legs of a dot represent the $\alpha_{1,2}$ and $\beta_{1,2}$ indices
in the tensor $g^m_{\alpha_1\al_2;\beta_1\bt_2}$.  The eight legs of a vertex
represent the internal indices in the rank-4 tensor
$T_{\alpha_1\al_2;\beta_1\bt_2;\gamma_1\ga_2;\lambda_1\la_2}$.  The indices on
the connected links are summed over which define the tensor-trace tTr.
} \label{Z2Z2D}
\end{figure}

In order to calculate the energy expectation values in these states
(and also to pave the way for generalizations) it is
convenient to write the trial wave function $\ket{\Psi_w}$ as a tensor
product state:
\begin{eqnarray}
\label{Z2wavefunction}
\ket{\Psi_{Z_2}}=\sum_{m_1,m_2,...}\tx{tTr}[\otimes_v T \otimes_l
g^{m_l}] |m_1,m_2,...\rangle  ,
\end{eqnarray}
where $m_l=0,1$ labels the up-spin state and the down-spin state on
link-$l$. To define the tensor-trace (tTr), one can introduce a
graphical representation of the tensors (see Fig. \ref{Z2Z2D}a).  Then tTr
means summing over all indices on the connected links of
tensor-network. The $Z_2$ string-net condensed ground state that we
discussed above is given by the following choice of tensors:
\begin{eqnarray}
\label{T}
T_{\alpha\beta\gamma\delta}=\left\{
\begin{array}{cc}  
1, &
{\rm{if}} \quad \alpha+\beta+\gamma+\delta \quad \rm{even} \\ 
0,& {\rm{if}} \quad \alpha+\beta+\gamma+\delta \quad \rm{odd}
\end{array}\right.
\end{eqnarray}
\begin{eqnarray}
\label{g}
g^0_{00}=1, \ \ \ \ g^1_{11}=w, \ \ \ \ \text{
others}=0,
\end{eqnarray}
with internal indices like $\alpha$ running over $0,1$. The rank-3
tensor $g$ behaves like a projector which essentially set the
internal index equal to the physical index so that $\alpha=1$
represents a string and $\alpha=0$ represents no string. The meaning
of the tensor $T_{\alpha\beta\gamma\delta}$ is also clear, it just
enforces the closed string constraint, only allowing an even number
of strings to meet at a vertex.

\begin{figure}
\begin{center}
\includegraphics[scale=0.6]
{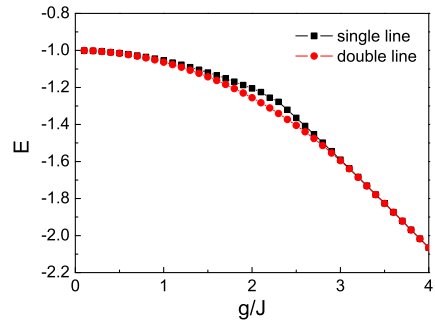}
\end{center}
\caption{
(Color online) The black squares are average energies of the $Z_2$ model
\eqn{HZ2} for the single-line tensor network \eqn{T} and \eqn{g}.  The red
dots are average energies for the double-line tensor network \eqn{TD} and
\eqn{gD}.
} \label{Z2energy}
\end{figure}

Once we have expressed the trial wave function as a TPS, we can use
the TERG method\cite{GLWtergV} to calculate the
average energy in a very efficient way.\footnote{We can absorb the
$g$-tensor into the $T$-tensor when we construct the double-tensor
in the TERG calculation.} The resulting average energy as a function
of $g/J$ is plotted in Fig.  \ref{Z2energy}.  From the discontinuity
in the slope, we see that there is a first order phase transition at
$g/J\approx 2.3$ between the two states with and without string-net
condensation.

How good is this result? On a quantitative level, it is not very
good: the phase transition is known to occur at $(g/J)_c \approx 3.044$.\cite{BD0210}
However, this not surprising since we used the simplest possible
variational wave function. We expect the estimate for $(g/J)_c$ to
improve when we increase the number of variational parameters - for
example, by considering more general tensors $g$, $T$.

\begin{figure}
\begin{center}
\includegraphics[width=2.5in] {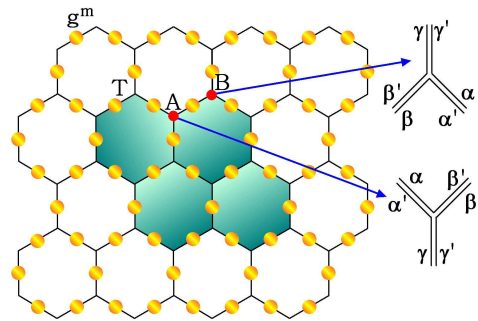}
\end{center}
\caption{The double-semion model on the honeycomb lattice. The
ground state wave function (\ref{semWF}) has a TPS representation
given by the above tensor-network.  Note that $T$ and $g$ has a
double-line structure. The vertices form a honeycomb lattice which
can be divided into A-sublattice and B-sublattice. } \label{semion}
\end{figure}

A more serious problem is that the result is wrong on a qualitative level:
the phase transition is known to be second order, not first order. This
problem cannot be overcome by blindly generalizing the tensors $g$, $T$.
Instead, we have to choose these tensors in a special way. To understand
the basic issue, let us consider another set of
variational tensors. In this scheme, the internal indices for the
$T$-tensors and $g$-tensors still run from $0$ to $1$, but each leg
now has two internal indices:\footnote{The expression
\eqn{TZ2} is valid only on the sublattice A.
The $T$ tensor has a slightly different form on the sublattice B.
For details see \eqn{semion1} and Fig. \ref{semion}.}
\begin{eqnarray}
\label{TZ2}
 T_{\al\bt\ga\la} &=& T_{\al_1\al_2;\bt_1\bt_2;\ga_1\ga_2;\la_1\la_2}
\nonumber\\
  &=& T^0_{\al_1\bt_1\ga_1\la_1}\delta_{\al_2\ga_1}\delta_{\ga_2\bt_1}\delta_{\bt_2\la_1}\delta_{\la_2\al_1}
\end{eqnarray}
with $\al_1,\al_2,\bt_1,\bt_2,\ga_1,\ga_2,\la_1,\la_2=0,1$ and
\begin{eqnarray}
\label{gD} g^0_{11,11}=g^0_{00,00}=1,\ \ g^1_{10,10}=g^1_{01,01}=1,\
\ \text{others}=0.
\end{eqnarray}
In such constructions, our tensors have a double-line structure (see
Fig.\ref{Z2Z2D}b). Again, $g^m$ are projectors that relate the internal indices
with physical indices. Here, on each leg of $g$-tensor and $T$-tensor, the
double line with the same value is projected to the spin-up state and the
double line with different values is projected to the spin-down state.

To maintain the $90$ degree rotational symmetry, we choose $T^0$ to
have a form (assuming $T^0_{1111}=1$):
\begin{eqnarray}
\label{TD}
x(1)&=&T^0_{0000}\nonumber\\
x(2)&=&T^0_{0111}=T^0_{1011}=T^0_{1101}=T^0_{1110}\nonumber\\
x(3)&=&T^0_{1000}=T^0_{0100}=T^0_{0010}=T^0_{0001}\nonumber\\
x(4)&=&T^0_{1100}=T^0_{0011}=T^0_{0110}=T^0_{1001}\nonumber\\
x(5)&=&T^0_{1010}=T^0_{0101}
\end{eqnarray}
We note that for such a choice of $T^0$ and $g^m$, the trial wave
function contain only closed string states.

\begin{figure}
\begin{center}
\includegraphics[scale=0.6]
{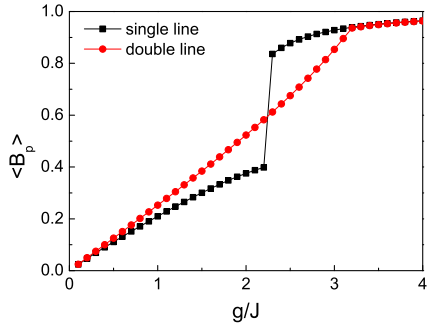}
\end{center}
\caption{(Color online)
$\langle B_p\rangle$ versus $g/J$. The
single-line variational wave function shows a jump in $\langle
B_p\rangle$, which indicates a first order phase transition around
$g/J=2.3$ (black squares). The double-line variational wave functions
has no jump in $\langle B_p\rangle$ but the discontinuity in the
derivative indicates a second order phase transition around
$g/J=3.2$ (red dots). } \label{Z2phase}
\end{figure}

Using the TERG approach to minimize the average energy, we find the
variational ground state energy which is plotted in Fig.
\ref{Z2energy}. We find that there is a phase transition between the
two phases with and without string-net condensation. But now the
phase transition is a second order phase transition at $g/J\approx
3.2$ (see Fig. \ref{Z2phase}). Note that this result is better than our
previous result both quantitatively and qualitatively.

The quantitative improvement is perhaps not surprising since we are
using more variational parameters. A more important issue is that the
double-line mean field theory correctly predicts a second order phase
transition, while the single line mean field theory did not. Why is this?

Note that there is a $Z_2$ redundancy in the double-line tensors
(like the gauge redundancy in gauge theory).
As we exchange values of $0$ and $1$ for all the internal indices of
the double-line tensors, we induce a $Z_2$ transformation on those
double-line tensors: $(T, g^m)\to (\t T,\t g^m)$.  However, such a
$Z_2$ transformation does not change the physical wave function: $
\tx{tTr}[\otimes_v T \otimes_l g^{m_l}] = \tx{tTr}[\otimes_v \t T
\otimes_l \t g^{m_l}] $.  Thus $(T,g^m)$ and $(\t T,\t g^m)$ are two
labels that label the same physical state.

The variational approach used here is similar to calculating an
average in a local classical statistical system.  The presence of a
$Z_2$ symmetry allows a classical system to have a $Z_2$ symmetry
breaking transition which is a second order phase transition. This
is the reason why the double-line tensors are capable of
producing a second order phase transition. In contrast, for the single-line
tensors, the corresponding classical system does not have
any symmetries, and as a result, it cannot describe a second order
transition.


We would like to mention that there is a duality transformation that
relate the 2D $Z_2$ gauge theory to transverse field Ising
model.\cite{K7959} Such a duality mapping allows us to relate the
phase transition between the deconfined and confined phases of the
$Z_2$ gauge theory to the spin ordered and disordered transition in
the transverse field Ising model. This is how we know that the
transition between the string-net condensed and non condensed phases
is a second order phase transition and that it occurs at critical
coupling $g/J \approx 3.044$. In fact, the double-line tensors exactly
realize the duality mapping between the 2D $Z_2$ gauge theory and
transverse field Ising model.  From the structure of the double
tensors in \eqn{gD} and \eqn{TD}, we see that each square loops in
Fig. \ref{Z2Z2D}b carries the same value of internal indices, which
correspond to the value of a dual spin (located at the center of the
square) in the dual Ising model. The string formed by the down-spins
on the links correspond to a domain wall in the dual Ising model.

\emph{Double-semion model:} Next we consider a more complicated
model where spins are located on the links of a honeycomb lattice
(see Fig. \ref{semion}):
\begin{align}
H &=  U\sum_{\v I} \Big(1- \prod_{\text{legs of } \v I} \si_{\v i}^{z} \Big)
-J\sum_{\v i}\si^z_{\v i}
\nonumber\\
&\ \ \ \
-g\sum_{\v p}(\prod_{\text{edges of }\v p} \si^x_{\v j})
(\prod_{\text{legs of }\v p} \imth^{\frac{1-\si^{z}_{\v j}}{2}}) ,
\label{sem}
\end{align}
where $\v i$ labels the links, $\v I$ labels the vertices and $\v p$ labels
hexagons.  Again we consider $U=\infty$ limit.
When $J=0$, the above model is exactly soluble and the exact ground
state is given by\cite{LWstrnet}
$|\Psi_\text{sem}\>=\sum_X (-)^{l(X)} |X\>$,
where $\sum_X$ sums over all the closed string configurations and $l(X)$ is
number of closed loops in $X$.  The ends of string in such a state have the
semion statistics.  When $g=0$, the model is also exactly soluble and the
spins all point up (\ie no strings) in the ground state.


To study the phase transition between the above two states,
again we choose the double-line tensors to construct
the trial wave function (see Fig. \ref{semion}).
The $T$-tensors in the vertices
are given schematically by
\begin{eqnarray}
\text{sublattice A}:\quad
T_{\alpha\alpha^\prime;\beta\beta^\prime;\gamma\gamma^\prime}&=&
T^A_{\alpha\beta\gamma}\delta_{\alpha\beta^\prime}
\delta_{\beta\gamma^\prime}\delta_{\gamma\alpha^\prime}\nonumber\\
\text{sublattice B}:\quad
T_{\alpha\alpha^\prime;\beta\beta^\prime;\gamma\gamma^\prime}&=&
T^B_{\alpha\beta\gamma}\delta_{\alpha^\prime\beta}
\delta_{\beta^\prime\gamma}\delta_{\gamma^\prime\alpha}\label{semion1}
\end{eqnarray}
where each internal index represented by one of the double lines
runs over $0,1$.  The tensor $T^A$ and $T^B$ is given by
\begin{eqnarray}
x(1)&=&T^A_{011}=T^A_{101}=T^A_{110} ;\quad x(2)=T^A_{001}=T^A_{100}=T^A_{010}\nonumber\\
x(3)&=&T^A_{111} ;\quad x(4)=T^A_{000}\nonumber\\
x(5)&=&T^B_{011}=T^B_{101}=T^B_{110} ;\quad x(6)=T^B_{001}=T^B_{100}=T^B_{010}\nonumber\\
x(7)&=&T^B_{111} ;\quad x(8)=T^B_{000}\label{semion2}
\end{eqnarray}
The $g$-tensors on the links are given by \eqn{gD}.
The trial wave function is obtained by summing
over all the internal indices on the connected links in the tensor
network (see Fig. \ref{semion}):
\begin{eqnarray}
\label{semWF}
\label{dsemion} | \Psi_\tx{dsemion}\rangle
=\sum_{\{m_l\}}\tx{tTr}[\otimes_v T \otimes_l g^{m_l}]
|m_1,m_2,...\rangle  .
\end{eqnarray}
Again, the physical indices and the internal indices have
a similar relation as in the $Z_2$ double-line tensors.
When $x(1)=x(5)=-\imth$, $x(2)=x(6)=\imth$ and $x(3)=x(4)=x(7)=x(8)=1$, the
tensor reproduces the right sign oscillations $(-)^{l(X)}$ essentially by
counting the number of left and right turns made by the string.

\begin{figure}
\begin{center}
\includegraphics[scale=0.6]
{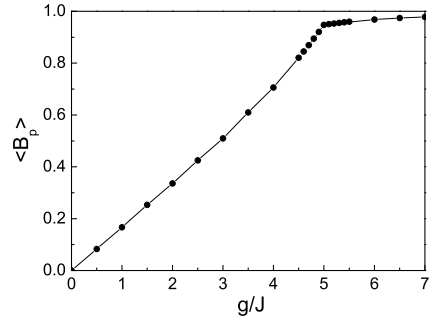}
\end{center}
\caption{$\langle B_p\rangle$ versus $g/J$, the discontinuity of the
derivative around $g/J=5.0$ indicates a second order
confinement-de-confinement phase transition.} \label{semionphase}
\end{figure}

We used the TERG approach to find the variational ground states for different
$g/J$.  Then we used the TERG approach to calculate $\<B_p\>$ for those
variational ground states.  The result is presented in Fig.
\ref{semionphase}. We see that there is a second order phase transition at
$g/J\approx 5.0$, which should correspond to the transition between the
string-net condensed and non condensed states. This agrees with the Monte
Carlo result where a second order phase transition appears at $(g/J)_c \approx
4.768$.\cite{BD0210} (Note that in the infinite $U$ limit, the above model is
equivalent to the usual $Z_2$ gauge model on honeycomb lattice, which is dual
to the transverse Ising model on triangle lattice.)

\emph{Conclusion:}
We have seen that the TERG approach is an effective way to
study topological phases and topological phase transitions, but one
needs to choose the variational tensors carefully. An important question
is how to choose the tensors in more general situations. One hint is that
all the string-net states constructed in \Ref{LWstrnet} can be expressed
naturally in terms of a generalization of the double-line tensor network,
which involves triple line tensors.\cite{GuString} This triple-line
tensor-network may correspond to the dual representation of the string-net
states and may correspond to a suitable choice for the variational TERG
approach. This may lead to a systematic variational approach for
topological phases and topological phase transitions.

\emph{Acknowledgements:}
This research is supported by the Foundational Questions Institute
(FQXi) and NSF Grant DMR-0706078.


\end{document}